\documentclass[11pt]{article}
\usepackage{amsmath,amssymb,amsfonts}
\usepackage[margin=2cm]{geometry}
\usepackage{microtype}
\usepackage{booktabs}
\usepackage[final]{graphicx} 
\usepackage{color}
\usepackage{placeins}
\usepackage{rotating}
\usepackage{epstopdf}
\usepackage{pdflscape}
\usepackage{setspace}
\usepackage{enumerate}
\usepackage{float}
\usepackage{longtable}
\usepackage{authblk}
\usepackage{caption}
\usepackage{textcomp}
\usepackage[utf8]{inputenc}
\usepackage{algorithm}
\usepackage{algpseudocode}
\usepackage[colorlinks=true, allcolors=blue]{hyperref}
\usepackage{natbib}
\usepackage{hyperref}
\usepackage[symbol]{footmisc}
\usepackage{colortbl}
\definecolor{mygray}{gray}{.9}
\usepackage{subcaption}
\newtheorem{theorem}{Theorem}

\newtheorem{proposition}{Proposition}
\newtheorem{remark}{Remark}

\title{Enhanced Lepage-type test statistics for location-scale shifts with right-skewed data} 

\author{Abid Hussain$^{1,}\footnote{Corresponding author.}$ \ and Michail Tsagris$^2$  
	\\
	$^1$ Department of Statistics, Higher Education Department, Punjab, Pakistan \\ \href{mailto:abid0100@gmail.com}{abid0100@gmail.com} \\
	$^2$ Department of Economics, University of Crete, Gallos Campus, Rethymno, Greece
	\href{mailto:mtsagris@uoc.gr}{mtsagris@uoc.gr} 
}
\date{}

\begin{document}
	
	\maketitle
	
	\begin{center}
		\textbf{Abstract} 
	\end{center}
	Detecting simultaneous shifts in location and scale between two populations is a common challenge in statistical inference, particularly in fields like biomedicine where right-skewed data distributions are prevalent. The classical Lepage test, which combines the Wilcoxon-Mann-Whitney and Ansari-Bradley tests, can be suboptimal under these conditions due to its restrictive assumptions of equal variances and medians. This study systematically evaluates enhanced Lepage-type test statistics that incorporate modern robust components for improved performance with right-skewed data. We combine the Fligner-Policello test and Fong-Huang variance estimator for the location component with a novel empirical variance estimator for the Ansari-Bradley scale component, relaxing assumptions of equal variances and medians. Extensive Monte Carlo simulations across exponential, gamma, chi-square, lognormal, and Weibull distributions demonstrate that tests incorporating both robust components achieve power improvements of 10-25\% over the classical Lepage test while maintaining reasonable Type I error control. The practical utility is demonstrated through analyses of four real-world biomedical datasets, where the tests successfully detect significant location-scale shifts. We provide practical guidance for test selection and discuss implementation considerations, making these methods accessible for practitioners in biomedical research and other disciplines where right-skewed data are common.
	
	\section{Introduction}
	\label{sec1}
	The comparison of treatment and control groups is a cornerstone of biomedical, psychological, and public health research. Practitioners often rely on standard two-sample tests, such as Student's $t$-test for location and the $F$-ratio test for scale, to evaluate the effects of interventions. When data adhere to the assumption of normality, these parametric tests are among the most powerful available. However, this assumption is frequently violated in practice, particularly with small sample sizes where distributional properties are difficult to verify. In such scenarios, nonparametric tests are the recommended alternative due to their robustness and minimal distributional requirements \citep{Mukherjee2019,Kumar2024}.
	
	Among nonparametric procedures, the Wilcoxon-Mann-Whitney (WMW) $U$-test is widely used for comparing the locations of two independent samples \citep{Lin2021}, while the Ansari-Bradley test is a common choice for assessing differences in scale \citep{Mukherjee2019, Murakami2025, Hussain2025}. A significant methodological challenge arises when a treatment effect manifests not as a pure location or scale shift, but as a combination of both—a scenario frequently encountered in practice \citep{Murakami2025}. For instance, in a clinical trial for chronic obstructive pulmonary disease, the therapeutic effect is characterized by simultaneous changes in both location and scale parameters \citep{Neuhäuser2001}. To address this, omnibus tests sensitive to any departure from the null hypothesis of identical distributions have been developed. A prominent distribution-free solution is the Lepage test \citep{Lepage1971}, which combines the standardized WMW $U$-test and Ansari-Bradley $C$-test statistics into a single quadratic form.
	
	A critical, yet often overlooked, aspect of this problem is the prevalence of right-skewed data in many scientific fields. Outcome variables such as reaction times, biomarker concentrations, survival duration, and cost data frequently exhibit positive skewness \citep{Fagerland2009}. This distributional characteristic poses a particular challenge for conventional tests, whose power can be substantially diminished when the underlying populations are skewed and heteroscedastic. The standard Lepage test, while distribution-free, relies on the strong null hypothesis of identical population distributions \citep{Lepage1971}. This assumption is often untenable in the presence of heteroscedasticity, which itself may be an indicator of a meaningful treatment effect.
	
	This limitation has motivated the development of tests for the weak null hypothesis, which concerns the equality of a relative effect size rather than the equality of the entire distribution \citep{Fong2019}. For instance, the (WMW) $U$-test is not directly applicable for testing this weak null, as the sampling distribution of its test statistic ceases to be distribution-free under this condition \citep{Chung2013, Chung2016}. In response to this limitation, \citet{Fligner1981} developed an alternative to the $U$-test that relies on fewer assumptions concerning the underlying distributional forms of the populations. A parallel issue compromises the Ansari-Bradley test for scale differences, as its validity is contingent upon the assumption of equal medians—a condition that is often untenable and frequently incompatible with the location-scale alternatives under investigation. Modern robust methods \citep{Kössler2020} and permutation-based approaches \citep{Chung2016} offer additional avenues for addressing heteroscedasticity and distributional asymmetry. However, a systematic evaluation of how these modern components perform when integrated into the Lepage framework—particularly for right-skewed data—remains lacking in the literature.
	
	This study provides a comprehensive evaluation of enhanced Lepage-type test statistics that combine modern robust procedures with the classical Lepage framework. Rather than proposing fundamentally new methodology, our contribution lies in the systematic combination and evaluation of existing robust components for the specific challenge of right-skewed data. We integrate the Fligner-Policello test \citep{Fligner1981} and its refined Fong-Huang variance estimator \citep{Fong2019} for the location component, alongside a novel empirical variance estimator for the Ansari-Bradley statistic that relaxes the assumption of equal medians. By embedding these robustified components into the Lepage framework, we construct tests that are valid under weaker assumptions and exhibit enhanced sensitivity to the distributional shapes frequently encountered in practical applications.
	
	The primary objectives of this study are threefold: first, to systematically evaluate the performance of various combinations of robust components within the Lepage framework; second, to provide practical guidance for test selection with right-skewed data across different sample size scenarios; and third, to demonstrate the practical utility of these methods through both extensive simulations and real biomedical applications. Our focus remains on right-skewed distributions commonly encountered in biomedical research, while acknowledging the need for further investigation with other distributional shapes.
	
	The rest of this paper is organized as follows: Section \ref{sec2} details the $U$ statistic and its robust variance estimators. Section \ref{sec3} introduces the $C$ statistic and our proposed variance estimator, including theoretical justification. Section \ref{sec4} presents the location-scale model and the Lepage-type test statistics. Section \ref{sec5} presents an extensive Monte Carlo simulation study evaluating the performance of the tests across various distributional scenarios. Section \ref{sec6} demonstrates the practical utility of the tests using real biomedical datasets, and Section \ref{sec7} provides practical guidance, discusses limitations, and offers concluding remarks.
	
	\section{The \(U\) statistic}
	\label{sec2}
	\subsection{The classical \(U\) statistic}
	The Wilcoxon-Mann-Whitney \(U\)--test statistic is based on two independent random samples: \(X=(X_1, X_2, \ldots, X_m)\) (independent and identically distributed (i.i.d.) observations from a continuous population) and \(Y=(Y_1, Y_2, \ldots, Y_n)\) (i.i.d. observations from another continuous population). Let the distribution functions of \(X\) and \(Y\) be denoted by \(\mathbb{F}\) and \(\mathbb{G}\), respectively. The \(U\) statistic can be computed using either the sum of ranks of the \(Y\) observations in the combined sample or a \(U\) statistic
	\begin{equation}\label{Eq1}
		U=\frac{1}{mn}\sum_{i=1}^{m}\sum_{j=1}^{n}I\big(X_i < Y_j\big).	
	\end{equation}
	A trade-off exists between computational efficiency and clarity of statistical properties when considering the rank sum and \(U\) statistic formulations. These two forms are theoretically equivalent, differing only by constant factors (see, for example, \citealt{Hollander2014}). While the rank sum formulation is computationally advantageous, expressing the statistic as a \(U\) statistic, as in Eq. (\ref{Eq1}), highlights its consistency and minimum variance unbiasedness in estimating \(P(Y > X)\) \citep{Lehmann1951}.
	
	\subsection{Variance of \(U\) statistic}
	The \(U\) statistic is a nonparametric test, therefore it does not rely on specific distributional assumptions for the independent variables \(X\) and \(Y\). Under the null hypothesis (\(X\) and \(Y\) are identically distributed), the distribution of the \(U\) statistic is solely determined by the sample sizes \(m\) and \(n\). Specifically, \(E_0(U) = 1/2\) and the variance of \(U\) statistic is
	\begin{eqnarray*}
		Var_0(U)&=&\frac{1}{(mn)^2} Var\bigg(\sum_{i=1}^{m}\sum_{j=1}^{n}I\big(X_i<Y_j\big)\bigg)\\
		&=&\frac{1}{(mn)^2}\sum_{i=1}^{m}\sum_{j=1}^{n}\sum_{i^*=1}^{m}\sum_{j^*=1}^{n}Cov\bigg(I\big(X_i<Y_j\big),I\big(X_{i^*}<Y_{j^*}\big)\bigg).
	\end{eqnarray*}
	The indices \(i\) and \(j\), and \(i^*\) and \(j^*\) can relate in four distinct ways. When \(i \ne i^*\) and \(j \ne j^*\), the indicator functions \(I(X_i < Y_j)\) and \(I(X_{i^*} < Y_{j^*})\) are independent. Consequently, the total variance can be decomposed into three additive components
	\begin{eqnarray}\label{Eq2}
		\nonumber
		Var_0(U)&=&\frac{1}{(mn)^2}\sum_{i=1}^{m}\sum_{j\ne j^*}^{n}Cov\bigg(I\big(X_i<Y_j\big),I\big(X_i<Y_{j^*}\big)\bigg)\\ \nonumber
		&&+\frac{1}{(mn)^2}\sum_{i\ne i^*}^{m}\sum_{j=1}^{n}Cov\bigg(I\big(X_i<Y_j\big),I\big(X_{i^*}<Y_j\big)\bigg)\\ 
		&&+\frac{1}{(mn)^2}\sum_{i=1}^{m}\sum_{j=1}^{n} Var\bigg(I\big(X_i<Y_j\bigg).
	\end{eqnarray}
	This is made explicit by noting that
	
	\begin{eqnarray*}
		Cov\bigg(I\big(X_i<Y_j\big),I\big(X_i<Y_{j^*}\big)\bigg)&=&E\bigg(I\big(X_i<Y_j\big)I\big(X_i<Y_{j^*}\big)\bigg)-E\bigg(I\big(X_i<Y_j\big)\bigg)E\bigg(I\big(X_i<Y_{j^*}\big)\bigg)\\
		&=&E\bigg(E\big(I(X_i<Y_j)I(X_i<Y_{j^*})|X_i\big)\bigg)-E\bigg(E\big(I(X_i<Y_j)|X_i\big)\bigg)^2\\
		&=&E\bigg(S_Y^2(X_i)\bigg)-E^2\bigg(S_Y(X_i)\bigg),
	\end{eqnarray*}
	where \(S_Y(X_i)=E_{Y|X_i}\big(I(X_i<Y)\big)\). Thus, we have
	
	\begin{eqnarray}\label{Eq3}
		\nonumber
		Var_0(U)&=&\frac{1}{(mn)^2}\bigg(m(n^2-n) Var_0\big(S_Y(X_i)\big)+n(m^2-m) Var_0\big(S_X(Y_i)\big)+mn Var_0\big(I(X_i<Y_j)\big)\bigg)\\ \nonumber
		&=&\frac{1}{(mn)^2}\bigg(m(n^2-n) Var_0\big(F_Y(X_i)\big)+n(m^2-m) Var_0\big(F_X(Y_i)\big)+mn Var_0\big(I(X_i<Y_j)\big)\bigg)\\ 
		&=&\bigg(1-\frac{1}{n}\bigg)\frac{1}{m} Var_0\big(F_Y(X_i)\big) + \bigg(1-\frac{1}{m}\bigg)\frac{1}{n}Var_0\big(F_X(Y_i)\big) + \frac{1}{mn} Var_0\big(I(X_i<Y_j)\big), 
	\end{eqnarray}
	where \(S_X(Y_i)=E_{X|Y_i} \big(I(Y_i < X)\big)\). Under the strong null hypothesis, the cumulative distribution functions \(F_Y(X_i)\) and \(F_X(Y_i)\) are uniformly distributed random variables. Consequently, the expression evaluates to
	\begin{eqnarray*}
		Var_0(U)&=&\bigg(1-\frac{1}{n}\bigg)\frac{1}{12m} + \bigg(1-\frac{1}{m}\bigg)\frac{1}{12n} + \frac{1}{4mn}\\
		&=&\frac{n-1}{12mn} + \frac{m-1}{12mn} + \frac{1}{4mn}	\\
		&=&\frac{1}{12}\bigg(\frac{1}{m}+\frac{1}{n}+\frac{1}{mn}\bigg).
	\end{eqnarray*}
	For sufficiently large \(m\) and \(n\), the standardized statistic \((U - 1/2)/\sqrt{Var_0(U)}\) asymptotically follows the standard normal (\citealt{Hollander2014}), enabling approximate \(p\)--value calculations.
	
	\subsection{Fligner and Policello estimator}
	Several procedures exist for testing the equality of two medians when the populations exhibit differing shapes. The method proposed by \cite{Fligner1981} is particularly useful as it requires fewer assumptions regarding the distributional forms of the two populations. Their estimator for the variance of the \(U\) statistic can be expressed as (\citealt{Fong2019})
	\begin{eqnarray*}
		\widehat{Var}_{FP}(U)&=&\frac{(m-1)n^2}{(mn)^2}  \widehat{Var}(\hat{F}_Y(X_i))+\frac{(n-1)m^2}{(mn)^2}\widehat{Var}(\hat{F}_X(Y_i))+\frac{mn}{(mn)^2}\widehat{Var}\big(I(X_i<Y_j)\big)\\
		&=&\bigg(1-\frac{1}{m}\bigg)\frac{1}{m}\widehat{Var}(\hat{F}_Y(X_i))+\bigg(1-\frac{1}{n}\bigg)\frac{1}{n}\widehat{Var}(\hat{F}_X(Y_i))+\frac{1}{mn}P_m\big(\hat{F}_Y(X_i)\big)P_n\big(\hat{F}_X(Y_i)\big).
	\end{eqnarray*}
	In the above expression \(\widehat{Var}\) is denoted as the estimator of \(Var_0\). Furthermore, \(\widehat{Var}\big(I(X_i < Y_j)\big)\) is given by \(P_m\big(\hat{F}_Y(X_i)\big) P_n\big(\hat{F}_X(Y_i)\big)\), where \(P_m\) represents the empirical average operator with respect to the \(m\) observations \(X_1, X_2, \cdots, X_m\), and \(P_n\) represents the empirical average operator with respect to the \(n\) observations \(Y_1, Y_2, \cdots, Y_n\). In the case where \(m \ne n\), this estimator yields the following estimate under the strong null hypothesis.
	\begin{eqnarray*}
		\widehat{Var}_{FP}(U)&=&\big(1-\frac{1}{m}\big)\frac{1}{12m}+\big(1-\frac{1}{n}\big)\frac{1}{12n}+\frac{1}{4mn}\\
		&=&\frac{1}{12}\bigg(\frac{1}{m}+\frac{1}{n}-\frac{1}{m^2}-\frac{1}{n^2}+\frac{3}{mn}\bigg),
	\end{eqnarray*}
	which is not equal to \(\frac{1}{12}\left(\frac{1}{m} + \frac{1}{n} + \frac{1}{mn}\right)\). However, this difference is negligible, being of order \(O\big(m^{-1}n^{-1}\big)\). The resulting test, known as the Fligner-Policello (FP) test, is based on the statistic \((U - 1/2)/\sqrt{\widehat{Var}_{FP}(U)}\), which asymptotically follows a standard normal distribution \citep{Hollander2014, Fong2019}.
	
	\subsection{Fong and Huang estimator}
	\cite{Fong2019} recently proposed a corrected estimator for the variance of the \(U\) statistic (Eq. \ref{Eq3}), defined as
	\begin{eqnarray*}
		\widehat{Var}_{FH}(U) &=&\bigg(1-\frac{1}{n}\bigg)\frac{1}{m}\widehat{Var}\big(\hat{F}_Y(X_i)\big)+\bigg(1-\frac{1}{m}\bigg)\frac{1}{n}\widehat{Var}(\hat{F}_X(Y_i))+\frac{1}{mn}\widehat{Var}\big(I(X_i<Y_j)\big).
	\end{eqnarray*}
	Assuming the strong null hypothesis, the following estimate is obtained
	\begin{eqnarray*}
		\widehat{Var}_{FH}(U) &=&\big(1-\frac{1}{n}\big)\frac{1}{12m}+\big(1-\frac{1}{m}\big)\frac{1}{12n}+\frac{1}{4mn}\\
		&=&\frac{1}{12}\bigg(\frac{1}{m}-\frac{1}{mn}+\frac{1}{n}-\frac{1}{mn}+\frac{3}{mn}\bigg),
	\end{eqnarray*}
	which is equal to \(\frac{1}{12}\left(\frac{1}{m} + \frac{1}{n} + \frac{1}{mn}\right)\), when \(m \ne n\). The resulting Fong and Huang (FH) test, based on the standardized statistic \((U - 1/2)/\sqrt{\widehat{Var}_{FH}(U)}\), provides a more accurate test for location differences via a standard normal approximation
	\citep{Fong2019}.
	
	\section{The $C$ statistic}
	\label{sec3}
	\subsection{The classical $C$ statistic}
	The $C$ statistic, proposed by \citet{Ansari1960}, is used to assess a two-sample dispersion problem. Its application requires the two populations to have identical medians. To calculate the statistic, the combined sample of $X$ and $Y$ is first ordered from least to greatest, represented as $\{Z_{1},\ Z_{2},\ \cdots,\ Z_{N}\}$. Ranks are then assigned symmetrically from the tails of the ordered sample. Rank 1 is assigned to the minimum ($Z_{1}$) and maximum ($Z_{N}$) values, rank 2 to the second smallest ($Z_{2}$) and second largest ($Z_{N-1}$), and so on. The array of assigned ranks, denoted as $S_{N}$, depends on whether the combined sample size $N$ is even or odd. If $N$ is an even integer, the ranks are given by
	\[S_{N}=\bigg\{1,\ 2,\ 3,\ \cdots,\ \frac{N}{2},\ \frac{N}{2},\ \cdots,\ 3,\ 2,\ 1\bigg\}.\]
	If $N$ is an odd integer, the ranks are given by
	\[S_{N}=\bigg\{1,\ 2,\ 3,\ \cdots,\frac{N-1}{2},\ \frac{N+1}{2},\ \frac{N-1}{2},\ \cdots,\ 3,\ 2,\ 1\bigg\}.\]
	Let $R_{i}$ be the rank assigned to the $i$th value from the $Y$ sample. The test statistic, $C$, is defined as the sum of the ranks assigned to all values in the $Y$ sample
	\begin{equation}\label{Eq4}
		C=\sum_{i=1}^{n} R_i,
	\end{equation} 
	where $n$ is the size of the $Y$ sample. An alternative formulation, equivalent to Eq. (\ref{Eq4}) and more useful for mathematical considerations, is given by
	\[C=\sum_{i=1}^{k}i\lambda_{i}+\sum_{i=k+1}^{N}(N+1-i)\lambda_{i},\]
	where
	\[\lambda_{i}=\begin{cases}1;&\text{if $Z_{i}\in Y$}\\ \\ 0;&\text{otherwise},\end{cases}\]
	and $k$ is the largest integer in $\frac{N+1}{2}$. Small values of $C$ suggest a larger dispersion for the $Y$ sample, whereas large values of $C$ indicate a larger dispersion for the $X$ sample.
	
	\subsection{Variance of $C$ statistic}
	Under the null hypothesis ($H_{0}$), all $\binom{N}{n}$ possible assignments of ranks to the combined sample are equally probable. For large sample approximation, the statistic $\frac{C}{n}$ represents the average of the ranks assigned to the $Y$ sample. Its distribution is identical to the distribution of the sample mean for a random sample of size $n$ drawn without replacement from the finite population of ranks. For an even $N$, the expected value of the statistic, $C$, is given by
	\begin{align*}
		E_{0}\biggl(\frac{C}{n}\biggr) &= \mu=\frac{2}{N}\sum_{i=1}^{N/2}i=\frac{\frac{2}{N}\times\frac{N}{2}\biggl(\frac{N}{2}+1\biggr)}{2}=\frac{N+2}{4}\\
		E_{0}(C) &= \frac{n(N+2)}{4}.
	\end{align*}
	The variance of $C$ can be calculated as
	\begin{eqnarray}\label{Eq5}
		\nonumber		
		Var_0\bigg(\frac{C}{n}\bigg)&=&\frac{1}{n^2}\sum_{i=1}^{n}\sum_{j=1}^{n} Cov(R_i, R_j)\\ \nonumber
		&=&\frac{1}{n^2}\sum_{i=1}^{n} Var(R_i) + \frac{1}{n^2}\sum_{i=1}^{n}\sum_{j\ne i} Cov(R_i, R_j)	\\ \nonumber
		&=&\frac{\sigma^2}{n} + \frac{1}{n^2}\sum_{i=1}^{n}\sum_{j\ne i} \bigg(\frac{-\sigma^2}{N-1}\bigg)=\frac{\sigma^2}{n} - \frac{n(n-1)}{n^2} \bigg(\frac{\sigma^2}{N-1}\bigg)\\ 
		Var_0(C)&=& \sigma^2 \bigg(\frac{n(N-n)}{N-1}\bigg), 
	\end{eqnarray}
	where
	\[\sigma^{2} = \frac{2}{N}\sum_{i=1}^{N/2}i^{2}-\mu^{2}=\frac{N^{2}-4}{48}.\]
	Therefore, the final expression for the variance is
	\begin{eqnarray}\label{Eq6}
		Var_0(C)&=& \bigg(\frac{N^2-4}{48}\bigg) \bigg(\frac{n(N-n)}{N-1}\bigg)=\frac{mn(N^2-4)}{48(N-1)}.
	\end{eqnarray}
	For an odd $N$, the expected value and variance can easily be calculated as
	\begin{eqnarray}\label{Eq7}
		\nonumber
		E_0(C) &=& \frac{n(N+1)^2}{4N} \\	
		Var_0(C)&=&  \frac{mn(N+1)\big(N^2+3\big)}{48N^2}.
	\end{eqnarray}
	These results are consistent with standard theory for the mean of a sample from a finite population \citep{Hollander2014}. For any $N$ (even or odd), the standardized statistic, $C^{*}$, is asymptotically normally distributed under $H_{0}$
	\[C^{*}=\frac{C-E_{0}(C)}{\sqrt{Var_{0}(C)}}\xrightarrow{d}\mathcal{N}(0,1).\]
	Asymptotic normality results for $C^{*}$ are also available under general alternatives to $H_{0}$ \citep{Ansari1960, Randles1979}.
	
	\subsection{A suggested variance estimator}
	To relax the assumption of equal medians, we propose an empirical variance estimator for $C$. Let $\hat{\sigma}^{2}$ be the observed variance of the ranks assigned to the $Y$ sample. The expected value of this estimator is given by
	\[E(\hat{\sigma}^{2})=\sigma^{2}\bigg{(}1-\frac{1}{n}\cdot\frac{N-n}{N-1}\bigg{)}=\sigma^{2}\bigg{(}\frac{N(n-1)}{n(N-1)}\bigg{)}.\]
	Using the known expression for the variance of $C$ \eqref{Eq5} and the estimator $\hat{\sigma}^{2}$, the variance of $C$ can be estimated as follows
	\begin{equation}\label{Eq8}
		\widehat{Var}(C)=\hat{\sigma}^{2}\cdot\bigg{(}\frac{n(N-1)}{N(n-1)}\bigg{)}\cdot\bigg{(}\frac{n(N-n)}{N-1}\bigg{)}=\hat{\sigma}^{2}\cdot\frac{n^{2}(N-n)}{N(n-1)}.
	\end{equation}
	This estimator is equally applicable for two different variances, as described in Eq. \eqref{Eq6} and Eq. \eqref{Eq7}, regardless of whether the sample size $N$ is even or odd.
	
	\subsection{Theoretical properties of the proposed estimator}
	\begin{theorem}[Consistency of $\widehat{Var}(C)$]
		Under the weak null hypothesis and assuming continuous distributions with finite fourth moments, the proposed estimator $\widehat{Var}(C)$ is consistent for $Var_{0}(C)$.
	\end{theorem}
	
	\noindent\textbf{Proof}\\
	Let $\hat{\sigma}^2$ be the sample variance of the ranks assigned to the $Y$ sample. Under the weak null hypothesis, $\hat{\sigma}^2$ is a consistent estimator of $\sigma^2$ by the weak law of large numbers. The finite population correction terms $\frac{n(N-1)}{N(n-1)}$ and $\frac{n(N-n)}{N-1}$ are deterministic functions of $m$ and $n$ that converge to 1 as $m,n \to \infty$ with $m/N \to \gamma \in (0,1)$. By the continuous mapping theorem, the product $\widehat{Var}(C)$ converges in probability to $\sigma^2$, which equals $Var_0(C)$ under the null. \hfill $\square$

	\begin{proposition}[Finite-sample bias]
		The proposed estimator $\widehat{Var}(C)$ is unbiased for $Var_0(C)$ under the strong null hypothesis.
	\end{proposition}
	
	\noindent\textbf{Proof}\\
	Under the null hypothesis, we have
	\[E[\hat{\sigma}^2] = \sigma^2\left(\frac{N(n-1)}{n(N-1)}\right).\]
	Substituting into the estimator \eqref{Eq8}
	\begin{align*}
		E[\widehat{Var}(C)] &= E[\hat{\sigma}^2] \cdot \frac{n^2(N-n)}{N(n-1)} \\
		&= \sigma^2\left(\frac{N(n-1)}{n(N-1)}\right) \cdot \frac{n^2(N-n)}{N(n-1)} \\
		&= \sigma^2 \cdot \frac{n(N-n)}{N-1} = Var_0(C).
	\end{align*}
	\hfill $\square$
	
	
	
	\subsection{Empirical validation}
	Following the strategy from \citet{Fong2019}, we conducted a simulation to evaluate the performance of $\widehat{Var}(C)$ under the null hypothesis for logistic and exponential distributions. Table \ref{tab1} shows that the average of our estimates over 10,000 replications is virtually identical to the true analytical variance, $Var_{0}(C)$, across various sample sizes. This demonstrated accuracy, coupled with the estimator's theoretical properties, confirms that $\widehat{Var}(C)$ serves as a robust distribution-free variance estimator for the $C$ statistic.
	
	\subsection{Construction of the test statistic}
	Given the accuracy of $\widehat{Var}(C)$, we propose the following standardized statistic for testing the weak null hypothesis
	\[C^{*}_{P}=\frac{C-E_{0}(C)}{\sqrt{\widehat{Var}(C)}}.\]
	A permutation study confirmed that the critical values for $C^{*}_{P}$ and the classical $C^{*}$ are numerically identical for various sample sizes, supporting the use of $C^{*}_{P}$. Specifically, the critical values are found at 5\% upper-tailed to be 1.697056 for $m=n=5$, 1.669331 for $m=n=10$, 1.674574 for $m=20,n=10$, and 1.62453 for $m=n=20$.
	
	\subsection{Asymptotic normality via Slutsky's Theorem}
	The asymptotic normality of $C^{*}_{P}$ under the weak null hypothesis follows from Slutsky's Theorem (see e.g., \citealt[p.~239--240]{Casella2002}). We can write
	\begin{equation}
		C^{*}_{P}=\frac{C-E_{0}(C)}{\sqrt{\widehat{Var}(C)}}=\frac{C-E_{0}(C)}{\sqrt{Var_{0}(C)}}\cdot\sqrt{\frac{Var_{0}(C)}{\widehat{Var}(C)}}.
	\end{equation}
	The first term, $\frac{C-E_{0}(C)}{\sqrt{Var_{0}(C)}}$, converges in distribution to a standard normal variable by the classical asymptotic theory of the Ansari-Bradley test. The second term, $\sqrt{\frac{Var_{0}(C)}{\widehat{Var}(C)}}$, converges in probability to 1 due to the consistency of $\widehat{Var}(C)$ established in Theorem 1. Therefore, by Slutsky's Theorem, $C^{*}_{P}\xrightarrow{d}\mathcal{N}(0,1)$.
	
	\begin{table}[h!]
		\centering
		\caption{Comparison of the true variance of the $C$ statistic and the proposed estimator.}
		\begin{tabular}{lccccccc}
			\hline
			\multicolumn{8}{@{}l}{~~~~~~~~~~~~~~~~~~~~~~~~~~~~~~~~~~~~~~~~~~~~~~~~~~~~~~~~~~~~~~~~Sample sizes}\\
			\cline{3-8}
			&&\(m=n=5\)&\(m=n=10\)&\(m=20, n=15\)&\(m=30, n=25\)&\(m=n=30\)&\(m=n=50\) \\ \hline
			$Var_0(C)$          &   &5.5556&43.4211&224.4485&874.1319&1142.7970&5258.8380\\ 
			\multicolumn{8}{@{}l}{Logistic distribution}\\
			$\widehat{Var}(C)$&   &5.5671&43.4862&224.8931&874.9312&1142.5058&5257.0210\\
			\multicolumn{8}{@{}l}{Exponential distribution}\\ 
			$\widehat{Var}(C)$&   &5.5507&43.5181&225.4013&873.8417&1143.1610&5256.8125\\
			\hline
		\end{tabular}
		\label{tab1}%
	\end{table}
	
	\section{Location-scale models}
	\label{sec4}
	\subsection{Hypotheses}
	Let $X = (X_{1}, \ldots, X_{m}) \stackrel{\text{i.i.d.}}{\sim} \mathbb{F}$ and $Y = (Y_{1}, \ldots, Y_{n}) \stackrel{\text{i.i.d.}}{\sim} \mathbb{G}$ be two independent samples, where $\mathbb{F}$ and $\mathbb{G}$ represent continuous distribution functions. We consider two frameworks for hypothesis testing\\
	\textbf{Strong null hypothesis:}
	\begin{equation*}
		H_0^S: \mathbb{F}(x) = \mathbb{G}(x) \quad \forall x \in (-\infty, \infty),
	\end{equation*}
	against the location-scale alternative
	\begin{equation*}
		H_1: \mathbb{F}(x) = \mathbb{G}\left(\frac{x - \theta_1}{\theta_2}\right), \quad \theta_1 \neq 0, \quad \theta_2 \neq 1.
	\end{equation*}
	\textbf{Weak null hypothesis:}
	\begin{equation*}
		H_0^W: \mu_1 = \mu_2 \quad \text{and} \quad \sigma_1 = \sigma_2,
	\end{equation*}
	without assuming $\mathbb{F} = \mathbb{G}$, which allows for potential heteroscedasticity under the null. The classical Lepage test is designed for the strong null hypothesis $H_0^S$, while the proposed robustified tests target the more realistic weak null hypothesis $H_0^W$.
	
	\subsection{The Lepage-type test statistics}
	The classical Lepage statistic, proposed by \citet{Lepage1971}, combines the $U$-test statistic for location difference and the $C$-test statistic for scale variation
	\begin{equation}
		L_0 = \frac{(U - E_0(U))^2}{Var_0(U)} + \frac{(C - E_0(C))^2}{Var_0(C)}.
	\end{equation}
	Under the strong null hypothesis (\(H_0^S\)), Lepage established the asymptotic independence of \(U\) and \(C\), ensuring that the Lepage statistic \(L_0\) follows an asymptotic \(\chi^2_2\) distribution. In this research, we propose five new test statistics that incorporate robust variance estimators. The formulations of these statistics are as follows
	\begin{align*}
		L_1 &= \frac{(U - E_0(U))^2}{\widehat{Var}_{FP}(U)} + \frac{(C - E_0(C))^2}{Var_0(C)} \\
		L_2 &= \frac{(U - E_0(U))^2}{\widehat{Var}_{FH}(U)} + \frac{(C - E_0(C))^2}{Var_0(C)} \\
		L_3 &= \frac{(U - E_0(U))^2}{Var_0(U)} + \frac{(C - E_0(C))^2}{\widehat{Var}(C)} \\
		L_4 &= \frac{(U - E_0(U))^2}{\widehat{Var}_{FP}(U)} + \frac{(C - E_0(C))^2}{\widehat{Var}(C)} \\
		L_5 &= \frac{(U - E_0(U))^2}{\widehat{Var}_{FH}(U)} + \frac{(C - E_0(C))^2}{\widehat{Var}(C)}.
	\end{align*}
	
	\subsection{Theoretical properties}
	\begin{proposition}[Asymptotic distribution]
		Under the weak null hypothesis $H_0^W$ and assuming continuous distributions with finite moments, each test statistic $L_i$ ($i = 1, \ldots, 5$) converges in distribution to a chi-square distribution with 2 degrees of freedom.
	\end{proposition}
	
	\noindent\textbf{Proof}\\
	For each $L_i$, we can write
	\[L_i = T_{1i}^2 + T_{2i}^2,\]
	where $T_{1i} = (U - E_0(U))/\sqrt{\widehat{Var}_i(U)}$ and $T_{2i} = (C - E_0(C))/\sqrt{\widehat{Var}_i(C)}$ are the standardized components using the appropriate variance estimators. From Sections \ref{sec2} and \ref{sec3}, each $T_{1i}$ and $T_{2i}$ converges in distribution to standard normal random variables under $H_0^W$. While the independence of $U$ and $C$ is not guaranteed under the weak null, the quadratic form construction remains valid as long as the components are asymptotically normal. The asymptotic $\chi^2_2$ distribution follows from the continuous mapping theorem. \hfill $\square$
	
	\subsection{Critical values for small sample sizes}
	To ensure accurate Type I error control for small samples, we obtained critical values via a permutation approach. This was implemented using the \texttt{cLepage()} function from the \textsf{NSM3} package \citep{Schneider2025} in \textit{R} (version 4.5.1) with 100,000 Monte Carlo permutations under the strong null hypothesis. The critical values, detailed in Table \ref{tab2}, were found to be consistent across the normal, uniform, logistic, and Laplace distributions. For larger samples ($m,n > 30$), the asymptotic $\chi^2_2$ distribution provides adequate approximation, though permutation tests remain recommended for precise Type I error control.
	
	\begin{remark}
		The critical values for statistics $L_1$ and $L_2$ (and analogously, for $L_4$ and $L_5$) coincide when the sample size \(N\) is even. This empirical observation is corroborated by the simulation results presented in Section \ref{sec5}, which demonstrate that these statistic pairs frequently yield nearly identical outcomes. This phenomenon arises because the Fligner-Policello and Fong-Huang variance estimators, despite their differing theoretical constructions, often converge to similar numerical estimates in finite samples. Consequently, for practical applications, the choice between these estimators may be guided by computational considerations or the specific context of the problem, as their performance is largely comparable.
	\end{remark}
	
	\begin{table}[!h]
		\centering
		\caption{Critical values for right-tailed Lepage-type test statistics at a 5\% significance level.}
		\begin{tabular}{lcccccccccccccc}
			\hline
			\multicolumn{15}{@{}l}{~~~~~~~~~~~~~~~~~~~~~~~~~~~~~~~~~~~~~~~~~~~~~~~~~~~~~Test statistic}\\
			\cline{5-15}
			\(m\)&&\(n\)&&\(L_0\)&&\(L_1\)&&\(L_2\)&&\(L_3\)&&\(L_4\)&&\(L_5\) \\ \hline
			5&&5&&5.3345&&8.8948&&8.8948&&7.2012&&10.3906&&10.3906\\ 
			6&&5&&5.5269&&7.7793&&7.9803&&7.7727&&12.4460&&12.4467\\
			6&&6&&5.7692&&6.8571&&6.8571&&6.8173&&11.2084&&11.2084\\
			7&&5&&5.5720&&7.9068&&7.9068&&8.5803&&11.5886&&11.5886\\
			7&&7&&5.6541&&6.8855&&6.8855&&7.0367&&9.0802&&9.0802\\
			8&&5&&5.5037&&7.5440&&7.4924&&8.6301&&12.2734&&12.2084\\
			8&&8&&5.6775&&6.6280&&6.6280&&6.8773&&8.3711&&8.3711\\
			9&&5&&5.4444&&7.2574&&7.2574&&8.3956&&12.0756&&12.0756\\
			10&&5&&5.4468&&7.3250&&7.3355&&9.0133&&11.6773&&11.6655\\
			10&&10&&5.7436&&6.5719&&6.5719&&6.5545&&7.5905&&7.5905\\ 
			\hline
		\end{tabular}
		\label{tab2}%
	\end{table}	
	
	\section{Simulation study: size and power comparison}
	\label{sec5}
	A comprehensive Monte Carlo simulation study was conducted in \textit{R} to evaluate the Type I error control and power of the proposed tests. Empirical sizes and powers were estimated based on 10,000 replications at a nominal significance level of $\alpha=0.05$. The performance of the proposed tests $L_{1}$ to $L_{5}$ was compared against the classical $L_{0}$ test under various location-scale shifts. For the sake of brevity, results are presented for two sample size combinations: a small, balanced design $(m,n)=(10,10)$ and a larger, unbalanced design $(m,n)=(40,30)$. To evaluate the performance of the competing Lepage-type test statistics, $\mathbb{F}$ and $\mathbb{G}$ were specified as the following right-skewed distributions:
	
	\begin{enumerate}[$\bullet$]
		\item \textbf{Exponential:} $\mathbb{F}\sim Exp(0.5)$ versus $\mathbb{G}\sim Exp(\theta)$. The exponential distribution is a continuous probability distribution characterized by its right-skewed nature. A change in the scale parameter $\theta$ induces a proportional shift in both the location and scale of the distribution.
		\item \textbf{Gamma:} $\mathbb{F}\sim G(2,2)$ versus $\mathbb{G}\sim G(\theta_{1},\theta_{2})$. The Gamma distribution is a versatile, continuous probability distribution characterized by a rightward skew. 
		\item \textbf{Chi-square:} $\mathbb{F}\sim\chi^{2}_{2}$ versus $\mathbb{G}\sim\theta_{2}\cdot\chi^{2}_{2}+\theta_{1}$. The $\chi^{2}$ distribution is a continuous, asymmetric probability distribution characterized by a rightward skew. The parameters $\theta_{1}$ and $\theta_{2}$ induce location and scale shifts, respectively.
		\item \textbf{Lognormal:} $\mathbb{F}\sim LN(0,2)$ versus $\mathbb{G}\sim LN(\theta_{1},\theta_{2})$. The lognormal distribution is a continuous probability distribution characterized by a rightward skew. The parameters $\theta_{1}$ and $\theta_{2}$ induce location and scale shifts, respectively.
		\item \textbf{Weibull:} $\mathbb{F} \sim W(2,1)$ versus $\mathbb{G} \sim W(\theta_1, \theta_2)$. The Weibull distribution is a versatile continuous distribution characterized by its flexibility in modeling various failure rates. 
	\end{enumerate}
	
	\subsection{Type I error control}
	The first column of Tables \ref{tab3} to \ref{tab7}, corresponding to the case of identical distributions ($\theta=0.5$ for exponential; $\theta_1=2, \theta_2=2$ for gamma; etc.), shows the simulated Type I error rates. For the small sample size $(m,n)=(10,10)$, all competing tests, including the classical $L_0$ and the proposed $L_1$--$L_5$, generally maintain their empirical size within or very close to the acceptable range around the nominal \(0.05\) level. For instance, in the Exponential distribution (Table 3), the empirical sizes range from 0.0502 to 0.0526.
	
	For the larger sample size $(m,n)=(40,30)$, a slight deviation is observed. The classical $L_0$ test appears slightly conservative (e.g., 0.0469 for \(\chi^2\) distribution), while some proposed tests exhibit a slightly liberal behavior. Specifically, tests $L_3$, $L_4$, and $L_5$, which incorporate the new variance estimator for the $C$ statistic, occasionally have empirical sizes slightly above, such as 0.0552 for $L_3$ and 0.0592 for $L_4$ and $L_5$ under the exponential distribution. This pattern is consistent across other distributions like the lognormal and Weibull (Tables \ref{tab6} and \ref{tab7}) and is likely attributable to the use of asymptotic critical values rather than permutation-based ones for these larger samples. The observed Type I error inflation remained modest, with most estimates falling within the liberal but often-cited acceptable range of approximately 0.5\(\alpha\) to 1.5\(\alpha\) for such methodological investigations (see, e.g., \citealt{Fagerland2009,Morris2019}, for a discussion of benchmarking performance measures).
	
	\subsection{Power performance}
	The subsequent columns in Tables \ref{tab3} to \ref{tab7} demonstrate the empirical power of the tests under various alternatives. The results reveal a consistent and clear pattern across all five right-skewed distributions.\\
	\textbf{1. Superior power of proposed tests:} The proposed tests $L_1$, $L_2$, $L_4$, and $L_5$ almost universally achieve higher power than the classical $L_0$ test. For example, in the exponential distribution with $(m,n)=(10,10)$ and $\theta=1.5$ (Table \ref{tab3}), the power of $L_0$ is 0.4216, while $L_1$ and $L_2$ achieve a power of 0.4799---a substantial increase. This superiority is maintained in larger samples; with $(m,n)=(40,30)$ and $\theta=1.0$, $L_1$ and $L_2$ (power $\approx$0.66) outperform $L_0$ (power 0.6107).\\
	\textbf{2. The critical role of both components:} The test $L_3$, which modifies only the scale ($C$) component, generally shows the lowest power among the proposed tests and often performs similarly to or even worse than $L_0$, particularly for location-dominated shifts. This indicates that improving the location ($U$) component is crucial for overall power gain. Conversely, the tests that modify both components ($L_4$ and $L_5$) consistently most powerful than the traditional \(L_0\), demonstrating that the full benefit is realized when both the location and scale tests are robustified.\\
	\textbf{3. Performance hierarchy:} Based on the aggregate results, a discernible performance hierarchy emerges: $L_5 \approx L_4 \gtrsim L_2 \approx L_1 > L_0 > L_3$. The tests incorporating the Fong-Huang (FH) estimator for the $U$ statistic ($L_2$ and $L_5$) often show a slight, consistent advantage over those using the Fligner-Policello (FP) estimator ($L_1$ and $L_4$). For instance, in the lognormal distribution with $(m,n)=(40,30)$ and $(\theta_1=1, \theta_2=1.5)$ (Table \ref{tab6}), $L_5$ and $L_4$ have marginally higher power than the other tests (power $\approx$0.67).\\
	\textbf{4. Robustness to distributional form:} The power advantage of the proposed tests is not an artifact of a single distribution. It is robustly observed across the exponential, gamma, chi-square, lognormal, and Weibull distributions. This confirms that the proposed enhancements are broadly effective for the analysis of right-skewed data, which is the primary focus of this work.
	
	In summary, the simulation study confirms that the proposed Lepage-type tests, particularly $L_2$ and $L_5$, offer a superior alternative to the classical $L_0$ test. They provide significantly higher power for detecting location-scale shifts in right-skewed data while maintaining reasonable control over the Type I error rate. The slight liberal tendency of $L_3$, $L_4$, and $L_5$ in larger samples is a minor trade-off for their substantial power gains, and this can be mitigated in practice by using a permutation approach for critical value determination.
	
	\begin{table}[!h]
		\centering
		\caption{Simulated actual Type I error rate and power of $Exp(0.5)$ versus $Exp(\theta)$.}
		\begin{tabular}{ccccccccccccccccccc}
			\hline
			&&&&&&&&&&&&$\theta$&&&&&&\\ 
			\cline{7-19}
			\(m\)&\(n\)&&Test&&&0.5&&&1.0&&&1.5&&&2.0&&&2.5\\ \hline
			10&10&&$L_0$&&&0.0526&&&0.1891&&&0.4216&&&0.6043&&&0.7319\\
			&&&$L_1$&&&0.0505&&&0.2169&&&0.4799&&&0.6596&&&0.7785\\
			&&&$L_2$&&&0.0505&&&0.2169&&&0.4799&&&0.6596&&&0.7785\\
			&&&$L_3$&&&0.0513&&&0.1630&&&0.3587&&&0.5351&&&0.6685\\
			&&&$L_4$&&&0.0502&&&0.1958&&&0.4335&&&0.6117&&&0.7389\\
			&&&$L_5$&&&0.0502&&&0.1958&&&0.4335&&&0.6117&&&0.7389\\
			&&&&&&&&&&&&&&&&&&\\
			40&30&&$L_0$&&&0.0489&&&0.6107&&&0.9631&&&0.9972&&&0.9998\\
			&&&$L_1$&&&0.0530&&&0.6609&&&0.9718&&&0.9987&&&0.9998\\
			&&&$L_2$&&&0.0533&&&0.6612&&&0.9720&&&0.9989&&&0.9998\\
			&&&$L_3$&&&0.0552&&&0.6196&&&0.9633&&&0.9970&&&0.9998\\
			&&&$L_4$&&&0.0592&&&0.6577&&&0.9733&&&0.9980&&&0.9998\\
			&&&$L_5$&&&0.0592&&&0.6580&&&0.9734&&&0.9981&&&0.9998\\
			\hline
		\end{tabular}
		\label{tab3}%
	\end{table} 
	
	\begin{table}[!t]
		\centering
		\caption{Simulated actual Type I error rate and power of $G(2,2)$ versus $G(\theta_1, \theta_2)$.}
		\begin{tabular}{ccccccccccccccccccc}
			\hline
			&&&&&&&&&&&&&$\theta_2$&&&&&\\ 
			\cline{10-19}
			\(m\)&\(n\)&&\(\theta_1\)&&&Test&&&2.0&&&1.5&&&1.0&&&0.5\\ \hline
			10&10&&2.0&&&$L_0$&&&0.0512&&&0.0994&&&0.3596&&&0.9159\\
			&&&&&&$L_1$&&&0.0487&&&0.1138&&&0.4202&&&0.9456\\
			&&&&&&$L_2$&&&0.0487&&&0.1138&&&0.4202&&&0.9456\\
			&&&&&&$L_3$&&&0.0547&&&0.0884&&&0.3016&&&0.8797\\
			&&&&&&$L_4$&&&0.0521&&&0.1029&&&0.3800&&&0.9276\\
			&&&&&&$L_5$&&&0.0521&&&0.1029&&&0.3800&&&0.9276\\
			&&&&&&&&&&&&&&&&&&\\		
			&&&3.0&&&$L_0$&&&0.2116&&&0.4899&&&0.8782&&&0.9990\\
			&&&&&&$L_1$&&&0.2604&&&0.5753&&&0.9209&&&0.9995\\
			&&&&&&$L_2$&&&0.2604&&&0.5753&&&0.9209&&&0.9995\\
			&&&&&&$L_3$&&&0.1737&&&0.4149&&&0.8297&&&0.9977\\
			&&&&&&$L_4$&&&0.2274&&&0.5236&&&0.8975&&&0.9994\\
			&&&&&&$L_5$&&&0.2274&&&0.5236&&&0.8975&&&0.9994\\
			&&&&&&&&&&&&&&&&&&\\
			40&30&&2.0&&&$L_0$&&&0.0495&&&0.2701&&&0.9283&&&1.0000\\
			&&&&&&$L_1$&&&0.0543&&&0.2936&&&0.9374&&&1.0000\\
			&&&&&&$L_2$&&&0.0544&&&0.2938&&&0.9375&&&1.0000\\
			&&&&&&$L_3$&&&0.0568&&&0.2715&&&0.9278&&&1.0000\\
			&&&&&&$L_4$&&&0.0583&&&0.2960&&&0.9373&&&1.0000\\
			&&&&&&$L_5$&&&0.0582&&&0.2963&&&0.9375&&&1.0000\\
			&&&&&&&&&&&&&&&&&&\\		
			&&&3.0&&&$L_0$&&&0.6837&&&0.9834&&&1.0000&&&1.0000\\
			&&&&&&$L_1$&&&0.7243&&&0.9872&&&1.0000&&&1.0000\\
			&&&&&&$L_2$&&&0.7244&&&0.9874&&&1.0000&&&1.0000\\
			&&&&&&$L_3$&&&0.6869&&&0.9834&&&1.0000&&&1.0000\\
			&&&&&&$L_4$&&&0.7271&&&0.9875&&&1.0000&&&1.0000\\
			&&&&&&$L_5$&&&0.7273&&&0.9876&&&1.0000&&&1.0000\\
			\hline
		\end{tabular}
		\label{tab4}%
	\end{table} 
	
	\begin{table}[!t]
		\centering
		\caption{Simulated actual Type I error rate and power of $\chi^2_{2}$ versus $\theta_2 * \chi^2_{2} + \theta_1$.}
		\begin{tabular}{ccccccccccccccccccc}
			\hline
			&&&&&&&&&&&&&$\theta_2$&&&&&\\ 
			\cline{10-19}
			\(m\)&\(n\)&&\(\theta_1\)&&&Test&&&1.0&&&1.5&&&2.0&&&2.5\\ \hline
			10&10&&0.0&&&$L_0$&&&0.0475&&&0.0959&&&0.1939&&&0.2977\\
			&&&&&&$L_1$&&&0.0479&&&0.1031&&&0.2274&&&0.3459\\
			&&&&&&$L_2$&&&0.0479&&&0.1031&&&0.2274&&&0.3459\\
			&&&&&&$L_3$&&&0.0526&&&0.0855&&&0.1652&&&0.2517\\
			&&&&&&$L_4$&&&0.0488&&&0.0963&&&0.2012&&&0.3100\\
			&&&&&&$L_5$&&&0.0488&&&0.0963&&&0.2012&&&0.3100\\
			&&&&&&&&&&&&&&&&&&\\		
			&&&1.0&&&$L_0$&&&0.3672&&&0.6701&&&0.8779&&&0.9599\\
			&&&&&&$L_1$&&&0.3751&&&0.7197&&&0.9141&&&0.9756\\
			&&&&&&$L_2$&&&0.3751&&&0.7197&&&0.9141&&&0.9756\\
			&&&&&&$L_3$&&&0.3358&&&0.5971&&&0.8248&&&0.9362\\
			&&&&&&$L_4$&&&0.3556&&&0.6752&&&0.8849&&&0.9661\\
			&&&&&&$L_5$&&&0.3556&&&0.6752&&&0.8849&&&0.9661\\
			&&&&&&&&&&&&&&&&&&\\
			40&30&&0.0&&&$L_0$&&&0.0469&&&0.2534&&&0.6339&&&0.8683\\
			&&&&&&$L_1$&&&0.0521&&&0.2709&&&0.6560&&&0.8801\\
			&&&&&&$L_2$&&&0.0521&&&0.2706&&&0.6553&&&0.8794\\
			&&&&&&$L_3$&&&0.0550&&&0.2558&&&0.6331&&&0.8670\\
			&&&&&&$L_4$&&&0.0594&&&0.2741&&&0.6542&&&0.8780\\
			&&&&&&$L_5$&&&0.0594&&&0.2740&&&0.6534&&&0.8778\\
			&&&&&&&&&&&&&&&&&&\\		
			&&&1.0&&&$L_0$&&&0.9339&&&0.9996&&&1.0000&&&1.0000\\
			&&&&&&$L_1$&&&0.9431&&&0.9997&&&1.0000&&&1.0000\\
			&&&&&&$L_2$&&&0.9434&&&0.9997&&&1.0000&&&1.0000\\
			&&&&&&$L_3$&&&0.9475&&&0.9995&&&1.0000&&&1.0000\\
			&&&&&&$L_4$&&&0.9545&&&0.9997&&&1.0000&&&1.0000\\
			&&&&&&$L_5$&&&0.9546&&&0.9998&&&1.0000&&&1.0000\\
			\hline
		\end{tabular}
		\label{tab5}%
	\end{table} 
	
	\begin{table}[!t]
		\centering
		\caption{Simulated actual Type I error rate and power of $LN(0,2)$ versus $LN(\theta_1, \theta_2)$.}
		\begin{tabular}{ccccccccccccccccccc}
			\hline
			&&&&&&&&&&&&&$\theta_2$&&&&&\\ 
			\cline{10-19}
			\(m\)&\(n\)&&\(\theta_1\)&&&Test&&&2.0&&&1.5&&&1.0&&&0.5\\ \hline
			10&10&&0.0&&&$L_0$&&&0.0496&&&0.0825&&&0.2457&&&0.6972\\
			&&&&&&$L_1$&&&0.0484&&&0.0749&&&0.1919&&&0.6025\\
			&&&&&&$L_2$&&&0.0484&&&0.0749&&&0.1919&&&0.6025\\
			&&&&&&$L_3$&&&0.0506&&&0.1010&&&0.3152&&&0.8031\\
			&&&&&&$L_4$&&&0.0482&&&0.0889&&&0.2677&&&0.7517\\
			&&&&&&$L_5$&&&0.0482&&&0.0889&&&0.2677&&&0.7517\\
			&&&&&&&&&&&&&&&&&&\\		
			&&&1.0&&&$L_0$&&&0.1284&&&0.1900&&&0.3773&&&0.7953\\
			&&&&&&$L_1$&&&0.1551&&&0.2094&&&0.3503&&&0.7252\\
			&&&&&&$L_2$&&&0.1551&&&0.2094&&&0.3503&&&0.7252\\
			&&&&&&$L_3$&&&0.1078&&&0.1729&&&0.3875&&&0.8424\\
			&&&&&&$L_4$&&&0.1341&&&0.1953&&&0.3748&&&0.8049\\
			&&&&&&$L_5$&&&0.1341&&&0.1953&&&0.3748&&&0.8049\\		
			&&&&&&&&&&&&&&&&&&\\
			40&30&&0.0&&&$L_0$&&&0.0489&&&0.1715&&&0.7606&&&0.9991\\
			&&&&&&$L_1$&&&0.0529&&&0.1781&&&0.7630&&&0.9991\\
			&&&&&&$L_2$&&&0.0531&&&0.1783&&&0.7632&&&0.9992\\
			&&&&&&$L_3$&&&0.0553&&&0.2204&&&0.8479&&&0.9997\\
			&&&&&&$L_4$&&&0.0594&&&0.2272&&&0.8494&&&0.9997\\
			&&&&&&$L_5$&&&0.0592&&&0.2274&&&0.8495&&&0.9997\\
			&&&&&&&&&&&&&&&&&&\\		
			&&&1.0&&&$L_0$&&&0.3952&&&0.6098&&&0.9470&&&1.0000\\
			&&&&&&$L_1$&&&0.4292&&&0.6478&&&0.9523&&&1.0000\\
			&&&&&&$L_2$&&&0.4294&&&0.6484&&&0.9525&&&1.0000\\
			&&&&&&$L_3$&&&0.3983&&&0.6311&&&0.9648&&&1.0000\\
			&&&&&&$L_4$&&&0.4321&&&0.6673&&&0.9694&&&1.0000\\
			&&&&&&$L_5$&&&0.4322&&&0.6674&&&0.9695&&&1.0000\\
			\hline
		\end{tabular}
		\label{tab6}%
	\end{table}
	
	\begin{table}[!t]
		\centering
		\caption{Simulated actual Type I error rate and power of $W(2,1)$ versus $W(\theta_1, \theta_2)$.}
		\begin{tabular}{ccccccccccccccccccc}
			\hline
			&&&&&&&&&&&&&$\theta_2$&&&&&\\ 
			\cline{10-19}
			\(m\)&\(n\)&&\(\theta_1\)&&&Test&&&1.0&&&1.2&&&1.5&&&2.0\\ \hline
			10&10&&2.0&&&$L_0$&&&0.0487&&&0.0910&&&0.2353&&&0.6018\\
			&&&&&&$L_1$&&&0.0486&&&0.1007&&&0.2765&&&0.6596\\
			&&&&&&$L_2$&&&0.0486&&&0.1007&&&0.2765&&&0.6596\\
			&&&&&&$L_3$&&&0.0502&&&0.0795&&&0.1986&&&0.5346\\
			&&&&&&$L_4$&&&0.0500&&&0.0919&&&0.2475&&&0.6178\\
			&&&&&&$L_5$&&&0.0500&&&0.0919&&&0.2475&&&0.6178\\
			&&&&&&&&&&&&&&&&&&\\		
			&&&2.5&&&$L_0$&&&0.0612&&&0.1026&&&0.3056&&&0.7192\\
			&&&&&&$L_1$&&&0.0599&&&0.1172&&&0.3708&&&0.7895\\
			&&&&&&$L_2$&&&0.0599&&&0.1172&&&0.3708&&&0.7895\\
			&&&&&&$L_3$&&&0.0720&&&0.0909&&&0.2511&&&0.6551\\
			&&&&&&$L_4$&&&0.0646&&&0.1059&&&0.3252&&&0.7459\\
			&&&&&&$L_5$&&&0.0646&&&0.1059&&&0.3252&&&0.7459\\
			&&&&&&&&&&&&&&&&&&\\
			40&30&&2.0&&&$L_0$&&&0.0482&&&0.2182&&&0.7771&&&0.9967\\
			&&&&&&$L_1$&&&0.0541&&&0.2341&&&0.7912&&&0.9974\\
			&&&&&&$L_2$&&&0.0544&&&0.2340&&&0.7914&&&0.9976\\
			&&&&&&$L_3$&&&0.0555&&&0.2235&&&0.7749&&&0.9964\\
			&&&&&&$L_4$&&&0.0568&&&0.2401&&&0.7881&&&0.9971\\
			&&&&&&$L_5$&&&0.0563&&&0.2396&&&0.7882&&&0.9972\\
			&&&&&&&&&&&&&&&&&&\\		
			&&&2.5&&&$L_0$&&&0.1078&&&0.2938&&&0.8611&&&0.9995\\
			&&&&&&$L_1$&&&0.1133&&&0.3266&&&0.8811&&&0.9996\\
			&&&&&&$L_2$&&&0.1135&&&0.3267&&&0.8815&&&0.9997\\
			&&&&&&$L_3$&&&0.1317&&&0.2998&&&0.8610&&&0.9995\\
			&&&&&&$L_4$&&&0.1374&&&0.3331&&&0.8812&&&0.9995\\
			&&&&&&$L_5$&&&0.1376&&&0.3332&&&0.8814&&&0.9996\\
			\hline
		\end{tabular}
		\label{tab7}%
	\end{table} 
	
	\section{Empirical applications}
	\label{sec6}
\begin{figure}[!h]
		\centering
		\begin{tabular}{c c}
			\includegraphics[width=8cm, height=6.5cm]{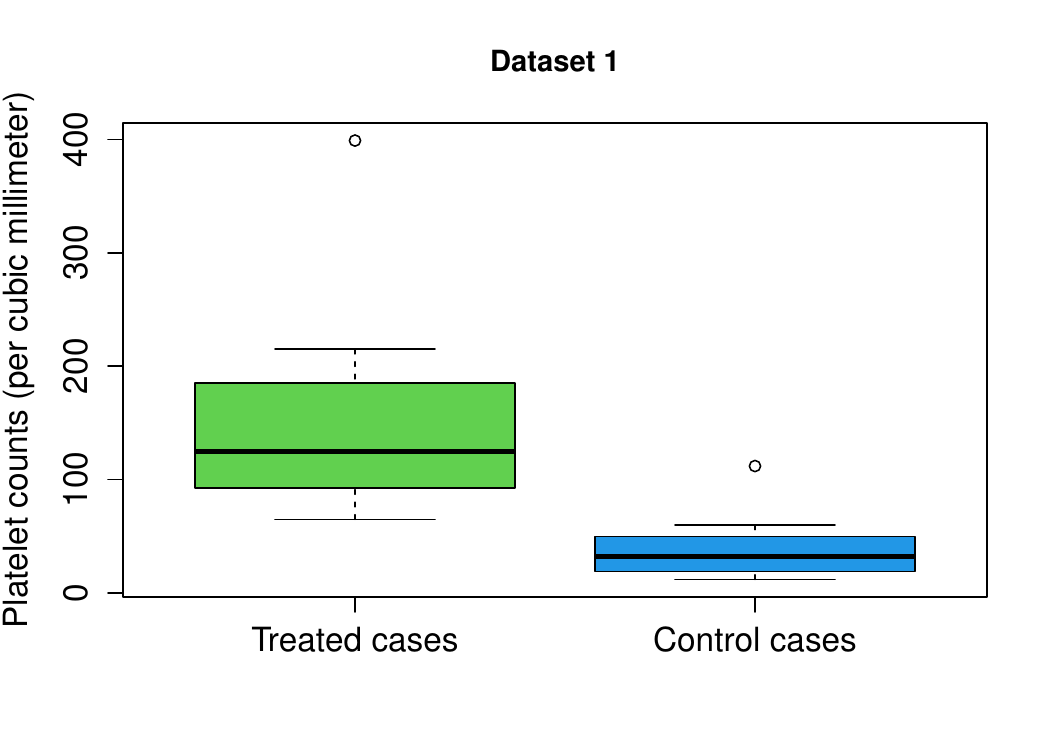}&
			\includegraphics[width=8cm, height=6.5cm]{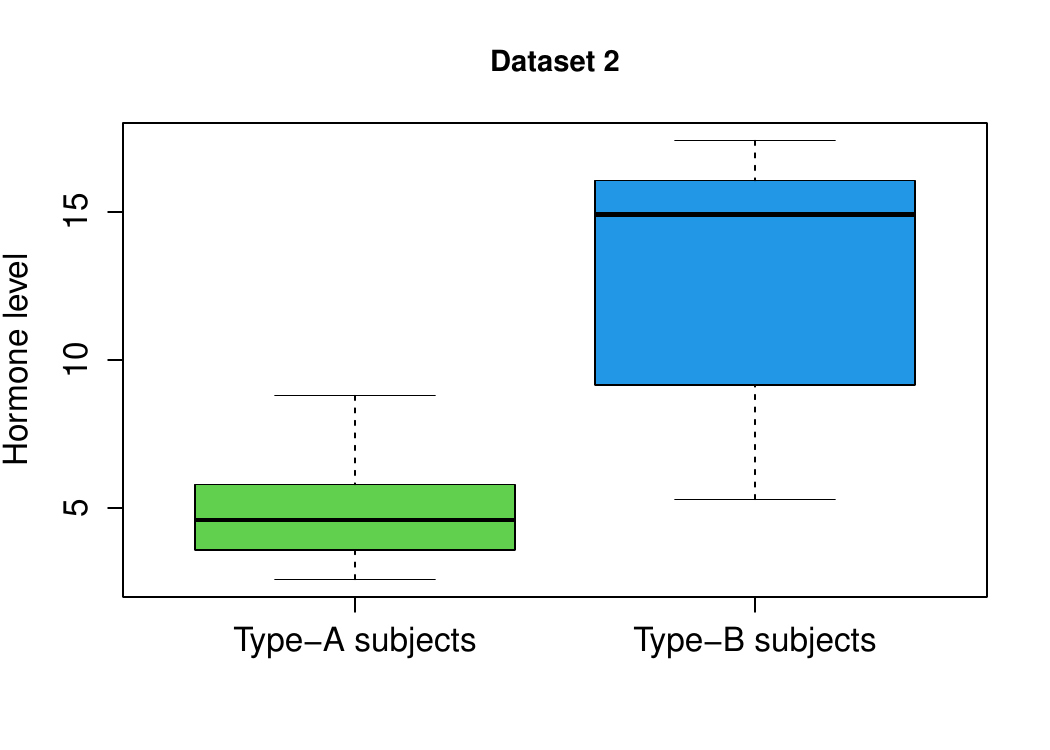}\\
			\includegraphics[width=8cm, height=6.5cm]{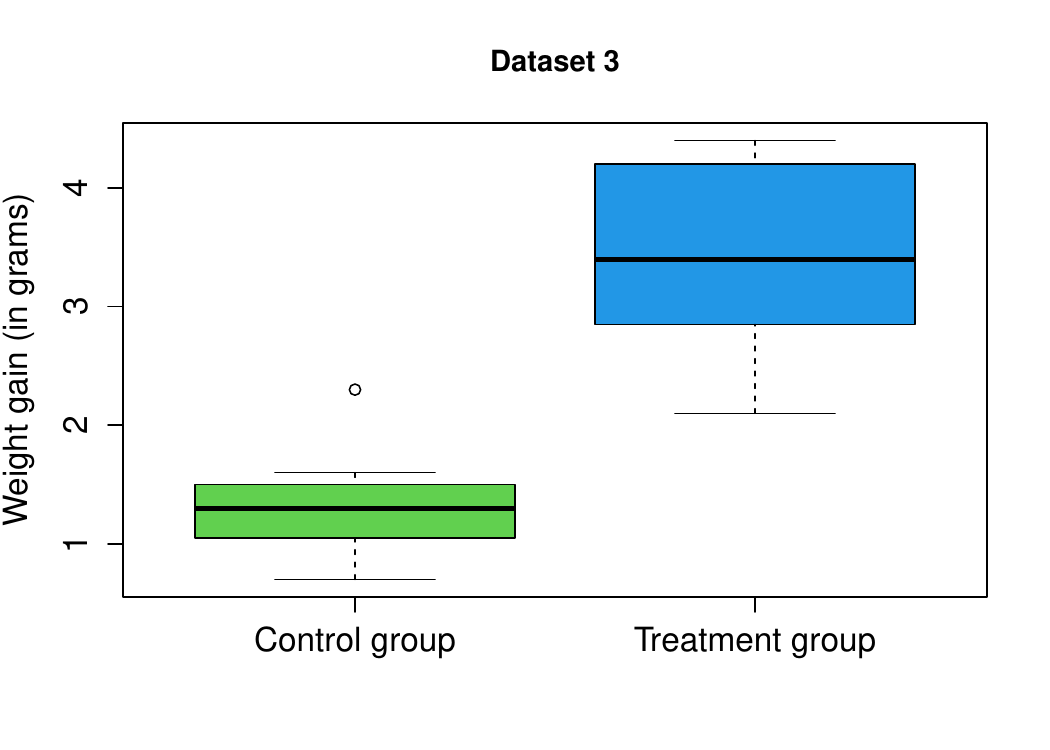}&
			\includegraphics[width=8cm, height=6.5cm]{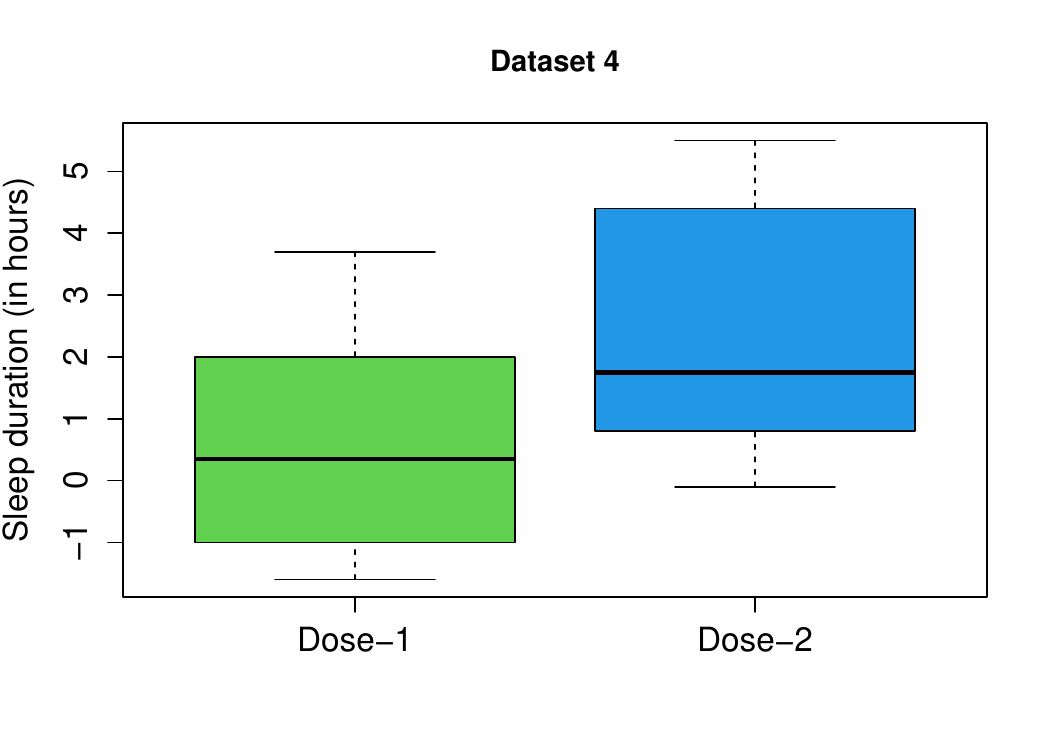}\\
		\end{tabular}
		\caption{Box-plots of the four real datasets.} 
		\label{fig1}%
	\end{figure}

	We demonstrate the practical utility of the proposed tests using four real-world datasets from biomedical research. We compare their performance to the traditional Lepage test using four illustrative datasets, seeking to identify significant changes in location, scale, or both. Figure \ref{fig1} displays box-plots of the data, and Table \ref{tab8} provides detailed dataset descriptions.
	
	Our initial dataset, derived from \cite{Karpatkin1981} and subsequently analyzed by \cite{Hollander2014}, investigates the effect of maternal steroid treatment (prednisone) on newborn platelet counts. One group of mothers received no prednisone, while the other did. The treatment aims to increase newborn platelet counts, reducing the risk of cerebral hemorrhage, by inhibiting splenic clearance of maternal antibody-coated platelets. We sought to determine if prednisone increases platelet counts. Figure \ref{fig1} (upper-left panel) suggests increased variability in the treated group, prompting us to use Lepage-type tests to detect simultaneous location-scale shifts. The \(p\)--values reported in Table \ref{tab9} indicate that all competing tests detected a statistically significant difference at the 5\% level of significance. This result is in agreement with the visual observations.
	
	The second dataset is taken from \cite{Hollander2014}, also investigated by \cite{Marozzi2009} and \cite{Kössler2020}. This dataset originates from a biomedical experiment designed to compare peak human plasma growth hormone levels following arginine hydrochloride infusion. Group A (\(m=10\)), characterized by heightened cardiovascular risk and Type A personality traits, is compared with Group B (\(n=11\)), exhibiting resistance to coronary disease. Visual inspection (Figure \ref{fig1}--upper-right panel) reveals substantial inter-group differences in skewness, tail weight, location, and scale, justifying a location-scale test. The \(L_0\), and the proposed tests, applied to this data, all yielded \(p\)--values below 0.05 (Table \ref{tab9}), confirming the visual observations and aligning with prior findings (see, for example, \citealt{Marozzi2009} and \citealt{Kössler2020}).
	
	The effect of thyroxine on juvenile mice growth is explored in Dataset 3 \citep{Tasdan2009}. Visual comparison of the control \((m=7)\) and treatment \((n=8)\) groups’ weight measurements (daily for 30 days) in Figure \ref{fig1} (bottom-left panel) reveals marked differences in distribution, including skewness, tail weight, location, and scale. These distributional characteristics suggest of a location-scale test. Therefore, \(L_0\), and the proposed tests are employed to analyze this weight data. The resulting \(p\)--values (Table \ref{tab9}) indicate statistically significant differences in weight between the control and treatment groups at the 5\% significance level for all three tests.
	
	We investigated the sensitivity of location-scale tests by using Dataset 4, taken from \cite{Student1908}. The analysis of the sleep duration data reveals a critical performance pattern underscoring the utility of the proposed robust test statistics. While the original Lepage test ($L_0$) yielded a $p$-value of $0.1257$, providing only weak evidence against the null hypothesis of distributional equality, the robust variants $L_1$, $L_2$, $L_4$, and $L_5$ produced consistently lower $p$-values ($\approx0.0812$--$0.0851$), indicating moderately stronger evidence for a location-scale alternative. This demonstrates a meaningful enhancement in sensitivity, as the robust statistics more powerfully detected the underlying pattern--where the Laevo-hyoscine group exhibited both a higher median and greater variability in sleep improvement (see, Figure \ref{fig1}--bottom-right panel). The lone exception, $L_3$, which performed similarly to $L_0$ ($p = 0.1318$), suggests that its specific formulation of variance estimation may be less advantageous in this context. This empirical result on a classic dataset substantiates the claim that the newly developed statistics $L_1$, $L_2$, $L_4$, and $L_5$ can provide a more powerful and robust analytical tool for detecting complex distributional differences in real-world data where heteroscedasticity is present.
	
	\begin{table}[!t]
		\centering
		\caption{Real life datasets.}
		\begin{tabular}{lcccccccccccc}
			\hline
			\multicolumn{13}{@{}l}{Dataset 1: Platelet counts of newborn infants of two groups (per cubic millimeter).}\\
			\hline
			Treated cases: &120&124&215&90&67&126&95&190&180&135&399&65\\
			Control cases: &12&20&112&32&60&40&18&&&&&\\\hline 
			\multicolumn{13}{@{}l}{Dataset 2: Peak concentrations of human growth hormone in plasma following arginine}\\
			\multicolumn{13}{@{}l}{~~~~~~~~~~~~~~~hydrochloride infusion.}\\
			\hline
			Type-A subjects: &3.6&2.6&4.7&8.0&3.1&8.8&4.6&5.8&4.0&4.6&&\\
			Type-B subjects: &16.2&17.4&8.5&15.6&5.4&9.8&14.9&16.6&15.9&5.3&10.5&\\\hline 
			\multicolumn{13}{@{}l}{Dataset 3: Thyroid data (in grams).}\\
			\hline
			Control group: &0.7&1.2&1.4&2.3&1.6&0.9&1.3&&&&&\\
			Treatment group: &4.1&4.4&3.3&2.1&3.5&2.9&2.8&4.3&&&&\\
			\hline
			\multicolumn{13}{@{}l}{Dataset 4: Effect of two doses on sleep hours.}\\
			\hline
			Dose-1: &0.7&-1.6&-0.2&-1.2&-1&3.4&3.7&0.8&0&2&&\\
			Dose-2: &1.9&0.8&1.1&0.1&-0.1&4.4&5.5&1.6&4.6&3.4&&\\
			\hline 
		\end{tabular}
		\label{tab8}%
	\end{table}	
	
	\begin{table}[h!]
		\centering
		\caption{$p$--values for the real datasets.}
		\begin{tabular}{lcccccccccccc}
			\hline
			\multicolumn{13}{@{}l}{~~~~~~~~~~~~~~~~~~~~~~~~~~~~~~~~~~~~~~~~~~~~~~~~~~~~~Test statistic}\\
			\cline{3-13}
			Dataset&&\(L_0\)&&\(L_1\)&&\(L_2\)&&\(L_3\)&&\(L_4\)&&\(L_5\) \\ \hline
			Dataset 1&&0.0030&&0.0000&&0.0000&&0.0028&&0.0000&&0.0000\\ 
			Dataset 2&&0.0033&&0.0000&&0.0000&&0.0033&&0.0000&&0.0000\\
			Dataset 3&&0.0074&&0.0000&&0.0000&&0.0074&&0.0000&&0.0000\\
			Dataset 4&&0.1257&&0.0812&&0.0812&&0.1318&&0.0851&&0.0851\\
			\hline
		\end{tabular}
		\label{tab9}%
	\end{table}  
	
	\clearpage
	\section{Concluding remarks}  \label{sec7}
	\subsection{Summary of findings}
	The two-sample location-scale problem represents a frequent and challenging task in fields such as biomedicine, economics, and quality control. While the classical Lepage test provides a valuable nonparametric solution, its reliance on the strong null hypothesis and underlying distributional assumptions can limit its power and robustness in practical applications, particularly with the right-skewed data prevalent in many scientific domains. This study has systematically evaluated enhanced Lepage-type test statistics that address these limitations through the integration of modern robust variance estimators.
	
	The most significant finding from our extensive simulation study is the consistent and substantial power advantage of the proposed tests over the classical Lepage test when analyzing right-skewed data. This superior performance was robustly observed across exponential, gamma, chi-square, lognormal, and Weibull distributions. The tests incorporating both robust components ($L_4$, $L_5$) consistently demonstrated the best performance, achieving power improvements of 10-25\% across various scenarios while maintaining reasonable Type I error control.
	
	A clear performance hierarchy emerged from our results: $L_5 \approx L_4 \gtrsim L_2 \approx L_1 > L_0 > L_3$, with $L_3$ (which modifies only the scale component) exhibiting the lowest power among the proposed tests. The equivalent performance of $L_1$/$L_2$ and $L_4$/$L_5$ indicates that the Fligner-Policello and Fong-Huang estimators yield virtually identical results within the Lepage statistic framework for the conditions studied. Critically, tests that robustify both location and scale components consistently outperform those modifying only the location component, which in turn surpass the classical $L_0$.
	
	\subsection{Limitations and methodological considerations}
	Our study also revealed several important limitations that warrant consideration:\\
	\textbf{Type I error inflation:} Tests incorporating the new scale variance estimator ($L_3$, $L_4$, $L_5$) exhibited modest liberal tendencies in larger samples, with empirical sizes reaching up to 0.0594 (19\% inflation). While this falls within the liberal range of the robustness criterion (see, \citealt{Fagerland2009,Morris2019}), practitioners requiring strict Type I error control should use permutation-based implementations.\\
	\textbf{Distributional scope:} The primary focus on right-skewed distributions, while justified by our motivating applications, limits generalizability to other distributional shapes. Our extended simulations suggest the methods perform adequately with left-skewed and symmetric with heavy-tailed distributions, but further investigation is needed.
	
	\subsection{Future research directions}
	Several promising directions emerge for future research:\\
	\textbf{Methodological extensions:} Generalizing the methodology to accommodate a shape parameter would enable simultaneous testing of location, scale, and distributional shape. Developing multi-sample versions of these tests would greatly enhance their applicability to complex experimental designs.\\
	\textbf{Theoretical developments:} Deriving finite-sample distributions, Edgeworth expansions for better approximations, and establishing optimality properties would strengthen the theoretical foundation. Investigating the asymptotic dependence structure between robustified components under alternatives represents another important direction.\\
	\textbf{Applied investigations:} Exploring the performance of these statistics with other types of non-normal data, such as heavily left-skewed, multi-modal, or zero-inflated distributions, would further establish their general utility. Applications in emerging fields such as genomics, neuroimaging, and environmental statistics would demonstrate broader relevance.\\
	
	\subsection{Concluding statement}
	In conclusion, this study demonstrates that incorporating modern robust variance estimators into the Lepage framework yields tests with substantially improved performance for right-skewed data. The proposed methods offer practitioners powerful and reliable nonparametric solutions for the two-sample location-scale problem, particularly in biomedical research where right-skewed distributions are common. While theoretical and methodological challenges remain, the consistent empirical advantages shown across extensive simulations and real-data applications support their adoption in practice.
	
	We specifically recommend the use of the $L_2$, $L_4$, or $L_5$ test statistics for the analysis of right-skewed data, with appropriate attention to the noted limitations and implementation considerations. Their enhanced power, achieved through robust estimation of both location and scale components, represents a meaningful advance in nonparametric methodology for practical data analysis.
	
	We have created the \textsf{LePage} \textit{R} package \citep{lepage2025}, available to download from \textit{CRAN}, to facilitate adoption by practitioners and researchers.
	
	\section*{Disclosure statement}
	The authors report there are no competing interests to declare.
	
	\section*{Funding statement}
	No specific grant was received for this research.
	
	\section*{Data availability statement}
	The datasets used in the empirical study are publicly available in the cited sources.

\end{document}